\documentstyle[12pt,epsf]{article}
\begin{document}
\baselineskip 16pt plus 2pt minus 2pt
\newcommand{\beq}{\begin{equation}}
\newcommand{\eeq}{\end{equation}}
\newcommand{\beqa}{\begin{eqnarray}}
\newcommand{\eeqa}{\end{eqnarray}}
\renewcommand{\thefootnote}{\#\arabic{footnote}}
\newcommand{\ve}{\varepsilon}
\newcommand{\eps}{\epsilon}
\newcommand{\krig}[1]{\stackrel{\circ}{#1}}
\newcommand{\barr}[1]{\not\mathrel #1}

\begin{titlepage}

%{\bf Draft,  \today}

\noindent REVISED VERSION \hfill TK 96 13

\hfill MKPH-T-96-05

\hfill hep--ph/9604294

\vspace{1.0cm}

\begin{center}

{\large { \bf Dispersion-theoretical analysis of the nucleon electromagnetic
form factors: Inclusion of time--like data\footnote{Work supported in
 part by the Deutsche Forschungsgemeinschaft (SFB 201)}}}

\vspace{1.2cm}

H.-W. Hammer$^{\ddag}$\footnote{Electronic address:
hammer@kph.uni-mainz.de},
Ulf-G. Mei\ss ner$^{\dag}$\footnote{Electronic address:
meissner@itkp.uni-bonn.de}\footnote{Address after Sept. 1$^{st}$, 1996:
FZ J\"ulich, IKP (Theorie), D-52425  J\"ulich, Germany},
D. Drechsel$^{\ddag}$\footnote{Electronic address:
drechsel@vkpmzp.kph.uni-mainz.de}

\vspace{0.8cm}

$^{\ddag}$Universit\"at Mainz, Institut f\"ur Kernphysik, J.-J.-Becher
Weg 45\\ D--55099 Mainz, Germany

\vspace{0.3cm}
$^{\dag}$Universit\"at Bonn, Institut f\"ur Theoretische Kernphysik, Nussallee
14-16\\ D--53115 Bonn, Germany
%\\and\\Institute for Nuclear Theory, University
%of Washington\\Seattle, WA 98195, USA

\vspace{0.4cm}

\end{center}

\vspace{1.5cm}

\begin{abstract}
We update a recent
dispersion--theoretical fit to the nucleon electromagnetic form
factors by including the existing data in the time--like region.
We show that while the time--like  data for the proton can be described
consistently with the existing world space--like data, this is not the
case for the neutron. Another measurement of the process $e^+ e^- \to 
\bar n n$ is called for. We furthermore sharpen the previous estimate
of the separation between the perturbative and the non--perturbative
regime, which is characterized by a scale parameter $\Lambda^2 
\simeq 10\,$GeV$^2$.
\end{abstract}

\vspace{2cm}

\vfill

\end{titlepage}

%\noindent {\bf 1.}
\section{Introduction}

A detailed understanding of the electromagnetic form factors of the nucleon
is not only of importance for revealing aspects of perturbative and
non--perturbative  nucleon structure but also serves as an important
ingredient for precise tests of the Standard Model like e.g. in the Lamb shift
measurements performed recently \cite{lamb}. These form factors have been
measured over a wide range of space--like momentum transfer
squared, $t = 0 \ldots -35\,$GeV$^2$
but also in the time--like region either in $\bar p p$ annihilation  or in
$e^+ e^- \to \bar p p, \bar n n$ collisions \cite{dataold}.
In particular, the FENICE experiment \cite{datfenice} has for the first time
measured the (magnetic) neutron form factor. 
These data and the corresponding ones for the
proton seem to indicate a new resonance at $\sqrt{t} = 1.85\,$GeV, which,
however, is not fully
consistent with the precise data from LEAR. For a comprehensive
summary, see Ref.\cite{baldini}. The data (space- and time--like) 
can be analyzed
in a largely model--independent fashion by means of dispersion relations.
In Ref.\cite{MMD} a new dispersion theoretical analysis of the nucleon
form factors has been performed based on the current world data basis for 
space--like momenta, i.e. for four--momentum transfer squared
$-t = Q^2 \ge 0$. This work improved upon the one of H\"ohler
et al. \cite{hoeh76} in various respects. These are the implementation
of the constraints from perturbative QCD (pQCD) at large momentum transfer,
the inclusion of the recent neutron--atom scattering length
determination  to constrain the neutron charge radius
and, of course, the inclusion of new data at low, moderate and high
momentum transfer (as listed in \cite{MMD}). However, the data for 
time--like momentum transfer ($t>0$) have not been accounted for.
In this paper, we will include these data and discuss the 
pertinent consequences. We stress that we do this without increasing the
number of vector meson poles compared to Ref.\cite{MMD} because only in that
way we can sharpen the analysis presented there.

%\bigskip
\section{Formalism}
%\noindent {\bf 2.}
Here, we briefly review the dispersion--theoretical formalism developed in 
Ref.\cite{MMD}. Assuming the validity of unsubtracted dispersion
relations for the four form factors $F_{1,2}^{(I=0,1)}(t)$, one separates the
spectral functions of the pertinent form factors into a hadronic
(meson pole) and a quark (pQCD) component as follows,
\beqa
F_i^{(I=0)} (t)& =& \tilde{F}_i^{(I=0)} (t) L(t) = \Bigg[ \sum_{I=0}
\frac{a_i^{(I=0)} \, L^{-1}(M^2_{(I=0)})}{M^2_{(I=0)} - t }\Bigg] \, \Bigg[
\ln \Bigg( \frac{\Lambda^2 - t}{Q_0^2} \Bigg)\Bigg]^{-\gamma}
 \\
F_i^{(I=1)} (t)& =& \tilde{F}_i^{(I=1)} (t) L(t) = \Bigg[
\tilde{F}_i^\rho (t) + \sum_{I=1}
\frac{a_i^{(I=1)} \, L^{-1}(M^2_{(I=1)})}{M^2_{(I=1)} - t }\Bigg] \, \Bigg[
\ln \Bigg( \frac{\Lambda^2 - t}{Q_0^2} \Bigg)\Bigg]^{-\gamma} \nonumber
\label{ffism} \eeqa
with
\begin{equation}
L(t) = \biggl[ \ln \bigg(\frac{\Lambda^2 - t}{Q_0^2} \biggr)
\biggr]^{-\gamma} \, \, \, ,
\label{Lt}
\end{equation}
and $F_i^\rho(t) = \tilde{F}_i^\rho (t)\, L(t)^{-1}$ parametrizes the two--pion
contribution (including the one from the $\rho$) in terms of the pion form
factor and the P--wave $\pi \pi \bar N N$partial wave amplitudes 
in a parameter--free manner (for details see \cite{MMD}).
Furthermore, $\Lambda$ separates the hadronic from the
quark contributions, $Q_0$ is related to $\Lambda_{\rm QCD}$
and $\gamma$ is the anomalous dimension,
\begin{equation}
F_i (t) \to 
(-t)^{-(i+1)} \, \biggl[ \ln\biggl(\frac{-t}{Q_0^2}\biggr)
\biggr]^{-\gamma} \, , \quad \gamma = 2 + \frac{4}{3\beta}
\, \, , \quad i = 1,2 \, \, ,
\label{fasy}
\end{equation}
for $t \to -\infty$ and  $\beta$ is the QCD $\beta$--function.
In fact, the fits performed in \cite{MMD} are rather
insensitive to the explicit form of the asymptotic form of the spectral
functions. To be specific, the additional factor
$L(t)$ in Eq.(\ref{ffism}) contributes to the spectral functions
for $t > \Lambda^2$, i.e. in some sense parametrizes the intermediate
states in the QCD regime, above the region of the vector mesons. The
particular logarithmic form has been chosen for convenience.
Obviously, the asymptotic behaviour of Eq.(\ref{Lt}) is obtained by
choosing the residues of the vector meson pole terms such that the
leading terms in the $1/t$--expansion cancel. In practice, the
additional logarithmic factor is of minor importance for the fit to
the existing data. 
The number of isoscalar and isovector poles in Eq.(\ref{ffism}) is determined
by the stability crtiterium discussed in detail in \cite{MMD}. In short,
we take the minimum number of poles necessary to fit the data. Specifically,
we have three isoscalar and three isovector poles.
The best fit to the available space--like proton {\it and}
neutron data will be  called fit 1 in what follows. Inclusion of the
time--like data leads to fit 2. We stress that we are keeping the number
of meson poles fixed so as not to wash out the predictive power. 
Due to the various constraints (unitarity, normalizations, superconvergence
relations) we end up having only three free parameters since two (three) of the
isovector (isoscalar) masses can be identified with the ones of 
physical particles, see below and \cite{MMD}.
We also have performed fits with more poles, these
will be mentioned but not discussed in detail below.

We should also make some remarks on the extraction of the time--like form
factors to be discussed. At the nucleon--anti-nucleon threshold, one has
\begin{equation}
G_M (4 m^2) = G_E (4 m^2) \quad ,
\label{Gthr}
\end{equation}
with $G_{E,M}$ the standard Sachs form factors
\begin{equation}
G_E = F_1 - \tau F_2 \, , \quad G_M = F_1 +F_2 \, , \quad \tau =
\frac{-t}{4m^2} \,\, \, ,
\label{sachs}
\end{equation}
and at large momentum transfer one expects the magnetic form factor to
dominate. From the data, one can not separate $|G_M|$ from $| G_E|$ so
one has to make an assumption, either setting $|G_M|= | G_E| = |G|$ or
$| G_E|= 0$. Most recent data are presented for the magnetic form factors
\cite{baldini}  and we will proceed accordingly, i.e. we fit 
the magnetic form factors in the time--like region.

\section{Results and discussion}

\noindent Consider first fit 1. The isoscalar masses are
$M_\omega=0.782\mbox{ GeV}$, $M_\phi=1.019\mbox{ GeV}$ and
$M_{S'}=1.60\mbox{ GeV}$ and the isovector ones are 
$M_{\rho'}=1.68\mbox{ GeV}$, $M_{\rho''}=1.45\mbox{ GeV}$ and
$M_{\rho'''}=1.69\mbox{ GeV}$. We note that apart from the $\rho'$ these
can be identified with masses of observed vector meson resonances. 
The corresponding residua are given in 
table~1, with the lower index $i=1,2$ referring to the vector and tensor
coupling, respectively, and the superscript $(n=1,2,3)$ enumerates the nth
pole. The isospin index $I=0,1$ is listed separately. 
The isovector residua
differ somewhat from the ones given in \cite{MMD} because we have updated
the data basis \cite{simon}\cite{nikhef}\cite{jourdan}. 
The corresponding numbers on the nucleon radii and 
vector-meson--nucleon coupling constants given there are, however, not
affected. The QCD parameters are $\gamma=2.148$, $\Lambda^2 = 9.73$~GeV$^2$
and $Q_0^2 = 0.35$~GeV$^2$. The consequences for the extraction of the
strange form factors are the same as in Ref.\cite{HMD}. 

\begin{center}

\renewcommand{\arraystretch}{1.5}
  
\begin{tabular}{|c||c|c|c|c|c|c|} \hline
%& & & & \\
& $a_1^{(1)}$ & $a_2^{(1)}$
& $a_1^{(2)}$ & $a_2^{(2)}$  & $a_1^{(3)}$ & $a_2^{(3)}$
\\ \hline
Fit 1, $I=0$& 0.747 & --0.122 & --0.737 & 0.162   & --0.039 & --0.041 \\
Fit 1, $I=1$& 189.3 & 35.90   & --3.158 & --6.476 & --186.7 & --30.92 \\
\hline
Fit 2, $I=0$& 0.752 & --0.121 & --0.752 & 0.159   & --0.020 & --0.038 \\
Fit 2, $I=1$& --9.913 & --4.731 & 13.01 & 0.263 & --3.497 & 2.947   \\
\hline \end{tabular}

\medskip

Table 1: Residues of the isoscalar ($I=0$) and the isovector ($I=1$)\\
 meson poles of the two fits discussed in the text.

\end{center}

The resulting Sachs form factors 
normalized to the canonical dipole fit
are displayed by the dashed lines in Fig.1a.\footnote{In case of the neutron
electric form factor we divide by $G_E^n (t)$ as given in Ref.\cite{pla}
for the Paris potential with the normalization readjusted to give the
exact value for the neutron charge radius.} 
Fit 1 has a $\chi^2$/datum of 1.06. 
It is worth mentioning that there is some inconsistency between the
new precise data from NIKHEF \cite{nikhef} and Mainz \cite{jourdan} 
on one side and the ones from Bonn \cite{bruins} on the other side.
The fit  prefers to stay slightly below the dipole approximation for
$Q^2 \le 1\,$GeV$^2$.

We now turn to fit 2. As stated before, we include the existing
time--like data but do not increase the number of meson poles. The number
of data points is increased by 30. In the fit, we keep the physical
masses fixed, i.e. we let the mass of the $\rho'$ free and
 allow  for variations in the pertinent residua. We find $M_{\rho'}
=1.40\mbox{ GeV}$, close to the mass of the $\rho''$. We also optimize
the fit by letting $\Lambda$ vary and find $\Lambda^2 = 12.0\,$GeV$^2$.
As noted in
\cite{MMD}, the third isovector mass always tends to come out close to
one of the other masses thus inducing some dipole--like structure.
The isovector residua are  mostly affected  as shown in table~1. 
The corresponding normalized Sachs form factors
for space--like $t$ are shown as solid lines in Fig.1a and the (magnetic)
time--like proton and neutron form factors are displayed in Fig.1b (again by
the solid lines). The $\chi^2$/datum of the fit increases to 1.47. This is
mostly due to the FENICE data for $|G_M^{p,n}(t)|$ for $t$ close to the
unphysical region at $t=4m^2$. In particular, we are not able to reproduce
the sharp increase of these data close to the unphysicsal region. We stress
that the resulting time--like $|G_M^p (t)|$ agrees nicely with 
(most of) the LEAR data.

\bigskip

\hskip 1in
\epsfxsize=4in
\epsffile{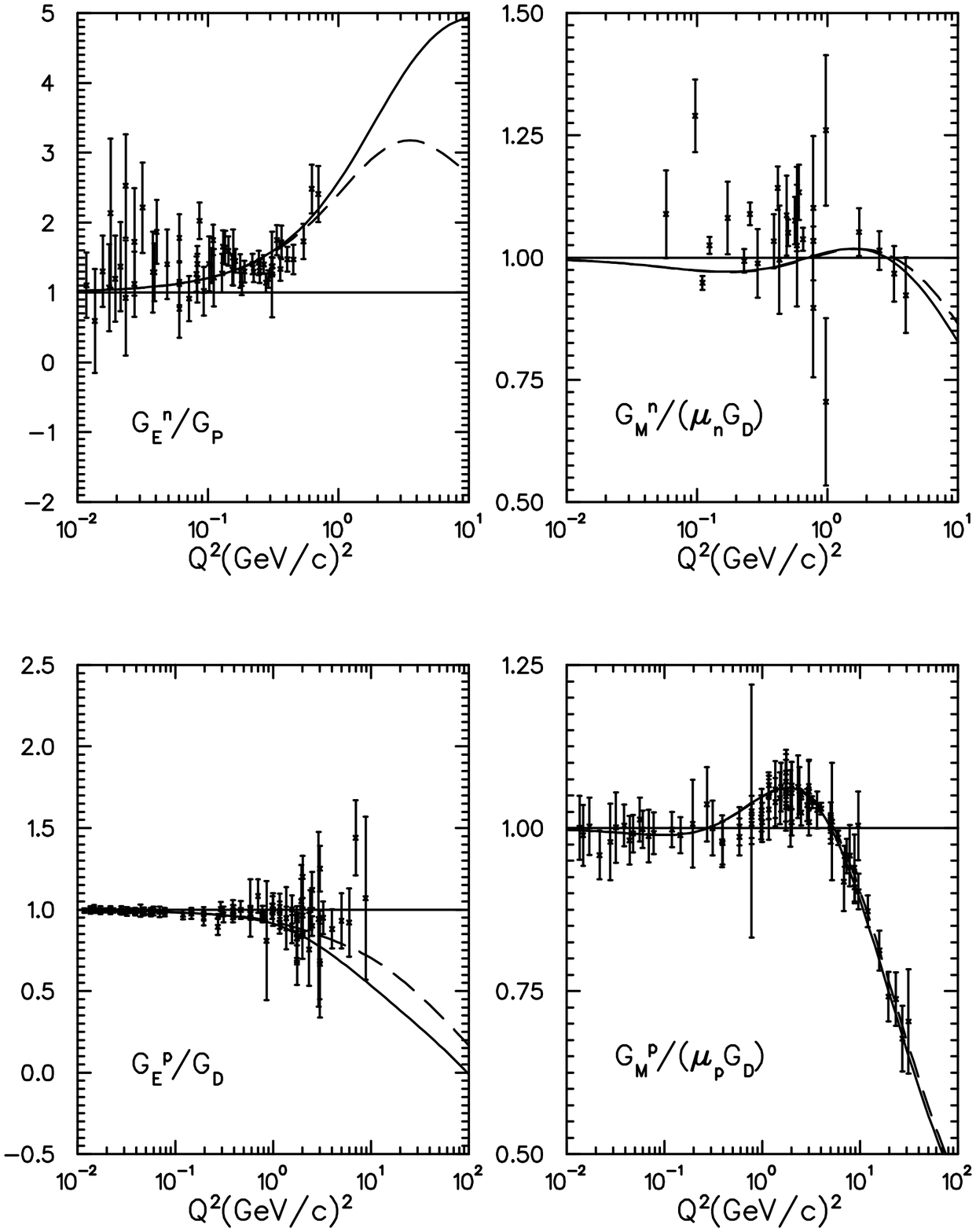}

%\bigskip

\noindent Fig.~1a:\quad  The space--like nucleon em form factors. 
Dashed lines: Fit 1, space--like data 

\qquad \quad $\,$ only. Solid lines: Fit 2, including also the time--like data.

%\smallskip}

\bigskip 

\noindent Since with a three pole fit in the isoscalar and isovector
channels we can not describe the FENICE data consistently, we have also
performed fits with more poles. For example, if one adds a fourth isoscalar
pole at the mass of the $\phi (1680)$, the fit does only improve marginally
as shown by the dashed lines in Fig.1b. The
$\chi^2$/datum of this fit is 1.46. Similarly, a fourth isoscalar pole
at 1.85 GeV does not lead to an improved fit.
Considering the present situation of
the data in the time--like region, in particular for the neutron, we
believe that first a more solid data basis is needed before 
one has to increase the
number of meson poles to get a better fit. This becomes more transparent
if one removes the FENICE neutron data from the data basis. Our standard three
pole fit then leads to better $\chi^2$/datum of 1.37. We also remark
that a smaller neutron time--like form factor would be more consistent with
pQCD estimates \cite{baldini}.
Some remarks on the sensitivity to the cut--off $\Lambda^2$ are in order.
While in \cite{MMD} one could vary its value between 5 and 15 GeV$^2$
without any drastic consequences for the fits, matters are differently here.
To keep the $\chi^2$/datum below 1.72, we can vary $\Lambda^2$ in the range 
from 9.3 to 16.3 GeV$^2$. We must stress, however, that there is 
still a sizeable contribution of the hadronic (pole) part of the
spectral functions at $Q^2 > \Lambda^2$, compare e.g. Fig.~7 in 
\cite{MMD}. This means that to test the predictions of pQCD one has to
go to momentum transfer  squared (much) larger than 20 GeV$^2$.    

\bigskip

\hskip 1.5in
\epsfxsize=3in
\epsffile{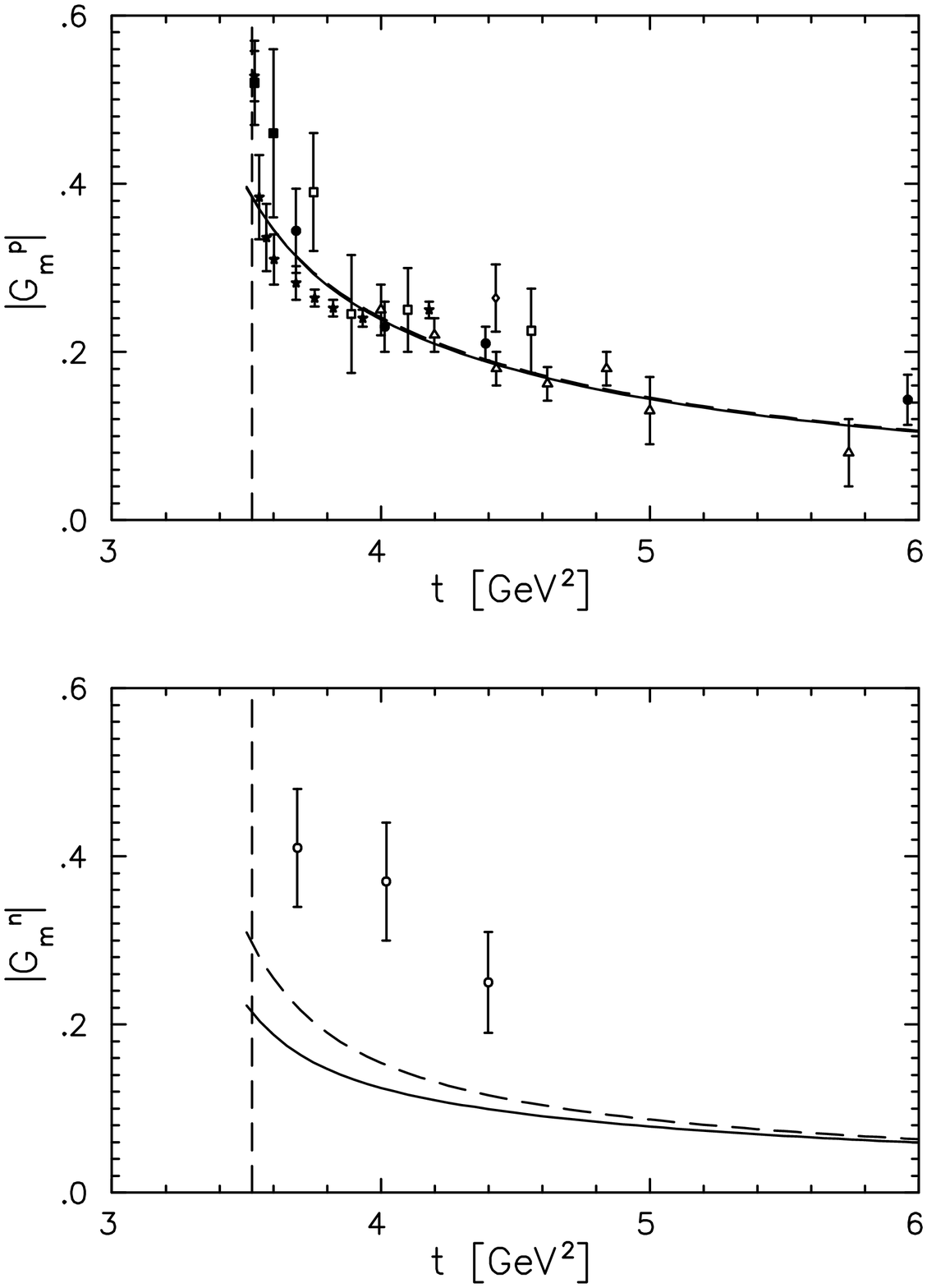}

\bigskip

\noindent Fig.~1b:\quad  The time--like nucleon em form factors. Solid
lines: Fit 2. Dashed lines: 

\qquad \quad $\,$ including in addition the $\phi (1680)$.

%\smallskip}

\bigskip \bigskip

To summarize, we have extended the dispersion--theoretical fit of
Ref.\cite{MMD} by including the existing data for time--like four--momentum
transfer. A fit with three isoscalar and three isovector poles does not
allow to consistently describe all the time--like data while still giving a
good description at space--like momentum transfer. Clearly,
the discrepancies in the time--like data for $|G_M^p (t)|$ need to be resolved
and additional experimental information on $|G_M^n (t)|$  is called for. 
Furthermore, a more refined treatment of the final--state interactions
in $e^+ e^- \to \bar n n$ might lead to a somewhat lower neutron form
factor in the time--like region.

\section*{Acknowledgements}
We are grateful to Gerhard H\"ohler for some instructive comments.
One of us (UGM) thanks the Institute for Nuclear Theory at the University
of Washington for its hospitality and the Department of Energy for partial
support during completion of this work.

\end{document}